\documentclass[a4paper,10pt]{article}
\makeatletter
  
  \@addtoreset{equation}{section}
\makeatother

\begin{document}

\setlength{\baselineskip}{16pt}
\setcounter{page}{0}
\thispagestyle{empty}

\vspace{10mm}
\begin{flushright}
{\large KEK-TH-919 \\
Sep. 2003 \hspace{4mm}$\;$\\}
\end{flushright}

\vspace*{60mm}

\begin{center}
{\LARGE  N=2 3d-Matrix Integral with Myers Term}\\

\vspace{20mm}
{\large Dan Tomino
\vspace{5mm}

{\it High Energy Accelerator Research Organization\\
Tsukuba, Ibaraki 305-0801, Japan\\

E-mail address : dan@post.kek.jp} }

\vspace{20mm}
{\bf Abstract}
\end{center}
An exact matrix integral is evaluated for a $2\times 2$ 3-dimensional matrix model with Myers term.
We derive weak and strong coupling expansions of the effective action.
We also calculate the expectation values of the quadratic and cubic operators.
Implications for non-commutative gauge theory on fuzzy sphere are discussed.

\newpage
\section{Introduction}
Matrix models in fuzzy homogeneous space \cite{K} can be thought as useful tools for further understanding of matrix dynamics, 
which is important from the viewpoint of non-perturbative definitions of superstring and M-theory \cite{IKKT}\cite{BFSS}.
Non-commutative (NC) gauge theories on homogeneous space are realized by these matrix models.
Especially a matrix model on fuzzy sphere: $SU(2)/U(1)$ is the simplest one of them.
It allows us to calculate the effective action of NC gauge theory explicitly and to determine the vacuum configuration of the space and gauge theory on it by minimization of the effective action.

The minimal model which realizes NC gauge theory on fuzzy sphere is a 3-dimensional matrix model with Myers term \cite{IKTW}:
\begin{eqnarray}
S=-\frac{1}{4}\mbox{Tr}[A_{\mu},A_{\nu}][A_{\mu},A_{\nu}]
+\frac{i}{3}\alpha \epsilon_{\mu\nu\rho}\mbox{Tr}[A_{\mu},A_{\nu}]A_{\rho}
+\frac{1}{2}\mbox{Tr}\bar{\Psi}\sigma^{\mu}[A_{\mu},\Psi].
\label{ac}
\end{eqnarray} 
where $A_{\mu=1,2,3}$ and $\Psi$ are $N\times N$ traceless Hermitian matrices.
$\Psi$ is two component Majorana spinor and $\sigma^{\mu}$ are Pauli matrices.
The equation of motion is 
\begin{eqnarray}
[A_{\mu},[A_{\mu},A_{\nu}]]+i\alpha\epsilon_{\mu\rho\nu}[A_{\mu},A_{\rho}]=0.
\label{EOM}
\end{eqnarray}
Fuzzy sphere is a non-trivial classical solution of (\ref{EOM})
\begin{eqnarray}
A_{\mu}=\alpha j_{\mu}\otimes 1_{n\times n},
\end{eqnarray}
where $j_\mu$ are spin $l$ representation of $SU(2)$.
Expanding the matrix model around this configuration, 
one can obtain NC gauge theories on fuzzy sphere.
The fixed parameters of this model are the matrix size $N=n(2l+1)$ and the constant $\alpha$.
In an analogous way, one can construct the NC gauge theories on fuzzy sphere by a deformed I{}IB matrix model: 
\begin{eqnarray}
S_{I{}IB}\rightarrow S_{I{}IB}+\frac{i}{3}\alpha\epsilon_{\mu\nu\rho}\mbox{Tr}[A_{\mu},A_{\nu}]A_{\rho}.
\end{eqnarray}

Recently effective actions of NC gauge theory on fuzzy sphere are investigated in \cite{IKTT} and \cite{IKTT2}.
Perturbative calculations have been carried out up to 2-loop level.
These calculations indicate an instability for the matrix models with Myers term and non-trivial realization of fuzzy $S_2\times S_2$ in I{}IB matrix model (without Myers term).
In strong coupling region of NC gauge theories, however, 2-loop calculations are less-reliable.
To further investigate strong coupling region, we need higher loop information, but loop calculations become harder and harder.
In this note, we try another approach: an exact calculation of a matrix model with Myers term.

Of course, an exact calculation of I{}IB matrix model with Myers term for any matrix size $N$ is very difficult.
However, we can perform an exact calculation of a 3d matrix model with Myers term when the size of matrices is 2.
The purpose of this paper is to report it.
$N\!\!=\!\!2$ 3d matrix model, of course, may be too simple.
But we hope to draw non-trivial information of strong coupling theories from this toy model.

The organisation of this paper is as follows.
We calculate the partition function of the $N\!\!=\!\!2$ 3d matrix model with Myers term exactly in section 2.
Without Myers term,  several exact calculations of matrix models have been reported \cite{ST}-\cite{refin7}.
The method of our calculation is modelled on \cite{ST}. 
We derive weak and strong coupling expansions of the partition function.
Next, we calculate the expectation value of some gauge invariant operators:
$\mbox{Tr} AA$ and $\mbox{Tr} \epsilon AAA$ in section 3 and 4.
Finally we discuss the results and their implication to NC gauge theory on fuzzy sphere.

\section{Partition Function}
We calculate the integral
\begin{eqnarray}
Z&=&\int\! dAd\Psi\;\exp(-S[A,\Psi]),
\label{defZ}
\end{eqnarray}
where $S$ is the matrix model action given by (\ref{ac}).

$A_{\mu}$ and $\Psi$ can be expanded by Pauli matrices
\begin{eqnarray}
A_{\mu}=\sum_{a=1}^3 A_{\mu}^a\frac{1}{2}\sigma^a, \qquad
\Psi=\sum_{a=1}^3 \Psi^a\frac{1}{2}\sigma^a.
\end{eqnarray} 

We fix the measure of the integral as follows
\begin{eqnarray}
dA&\equiv&\prod_{\mu=1}^3\prod_{a=1}^3\int_{-\infty}^{\infty}
\frac{dA_{\mu}^a}{\sqrt{2\pi}},\nonumber\\
d\Psi&\equiv&\prod_{\alpha=1}^2\prod_{a=1}^3\int\;d\Psi^a_{\alpha}.
\end{eqnarray} 

Using the $SO(3)$ rotational invariance, 
the matrix components $A_{\mu}^a$ can be taken as
\begin{eqnarray}
A^1&=&(L,0,0),\nonumber\\
A^2&=&(a_1,R_1\cos\theta_1,R_1\sin\theta_1),\\
A^3&=&(a_2,R_2\cos\theta_2,R_2\sin\theta_2) \nonumber.
\end{eqnarray}

In this parametrization 
\begin{eqnarray}
-\frac{1}{4}\mbox{Tr}[A_{\mu},A_{\nu}]^2
&=&\frac{1}{4}\left(
L^2(R_1^2+R_2^2)+R_1^2R_2^2\sin^2\theta+a_1^2R_2^2+a_2^2R_1^2
-2a_1a_2R_1R_2\cos\theta
\right),\nonumber\\
-\frac{i}{3}\epsilon_{\mu\nu\rho}\mbox{Tr}[A_{\mu}A_{\nu}]A_{\rho}
&=&LR_1R_2\sin\theta,
\end{eqnarray}
where $\theta=\theta_1-\theta_2$.

First, we carry out the fermion integral. The result is 
\begin{eqnarray}
\int\! dAd\Psi\;\exp(-S[A,\Psi])
=\int\!dA\;\mbox{Pf}(A)\exp(-S[A]),
\end{eqnarray}
where $\mbox{Pf(A)}$ is the Paffian:
\begin{eqnarray}
\mbox{Pf}(A)=-\frac{i}{12}\epsilon_{\mu\nu\rho}
\mbox{tr}[A_{\mu}A_{\nu}]A_{\rho}
\label{Paf}.
\end{eqnarray} 
The form of Paffian (\ref{Paf}) allows us to calculate $Z$ from the bosonic partition function ${\cal Z}$, i.e.
\begin{eqnarray}
Z&=&\frac{1}{4}\frac{\partial {\cal Z}}{\partial\alpha},\\
{\cal Z}&=&\int\! dA\;
\exp\left(\frac{1}{4}\mbox{Tr}[A_{\mu},A_{\nu}]^2
-\frac{i}{3}\alpha\epsilon_{\mu\nu\rho}\mbox{Tr}[A_{\mu}A_{\nu}]A_{\rho}\right).
\end{eqnarray}

Gathering the above results, the integral (\ref{defZ}) becomes
\begin{eqnarray}
Z&=&\frac{1}{4}\frac{\partial}{\partial\alpha}\frac{(2\pi)^2}{(\sqrt{2\pi})^9}\int^{\infty}_{-\infty}dL\int^{\infty}_{0}dR_1\int^{\infty}_{0}dR_2\int^{\infty}_{-\infty}da_1\int^{\infty}_{-\infty}da_2\int^{2\pi}_{0}d\theta\;
(L^2R_1R_2)\nonumber \\
&&\exp\Bigg[-\frac{1}{4}\left(
L^2(R_1^2+R_2^2)+R_1^2R_2^2\sin^2\theta+a_1^2R_2^2+a_2^2R_1^2
-2a_1a_2R_1R_2\cos\theta-4\alpha LR_1R_2\sin\theta
\right)\Bigg].
\nonumber \\
\label{intZ1}
\end{eqnarray}
After integrating over $a_1$ and $a_2$, (\ref{intZ1}) becomes
\begin{eqnarray}
&&\frac{1}{4}\frac{\partial}{\partial \alpha}\;\frac{4\pi}{(2\pi)^{9/2}}
\int^{\infty}_{-\infty}dL\int^{\infty}_{0}dR
\int^{2\pi}_{0}\frac{d\theta}{|\sin\theta|}\;
\frac{L^2}{\sqrt{L^2+R^2\sin^2\theta}}
\nonumber \\
&&
\left(\int^{\infty}_{-\frac{2\alpha LR\sin\theta}{\sqrt{L^2+R^2\sin^2\theta}}}dr\;e^{-\frac{1}{4}r^2}\;\right)
\exp\Bigg[-\frac{1}{4}L^2R^2\left(
1-\frac{4\alpha^2\sin^2\theta}{L^2+R^2\sin^2\theta}
\right)\Bigg].
\label{intZ2}
\end{eqnarray}
The $r$-integration in (\ref{intZ2}) can be done by using the Taylor expansion,
\begin{eqnarray}
\!\!\int^{\infty}_{-\frac{2\alpha LR\sin\theta}{\sqrt{L^2+R^2\sin^2\theta}}}\;e^{-\frac{1}{4}r^2}\;dr=
\left\{\sqrt{4\pi}
-\sum_{n=0}^{\infty}\frac{1}{(2n+1)n!}\left(\frac{-1}{4}\right)^n
\left(\frac{-2\alpha LR\sin\theta}{\sqrt{L^2+R^2\sin^2\theta}}\right)^{2n+1}\right\}\!\!.
\label{intZ3}
\end{eqnarray}
The odd power series for $L$ vanishes by the $L$-integral,
then only $\sqrt{4\pi}$ part in (\ref{intZ3}) is left. 
After changing the integration variables: $R=x/L$ and $L^2=yx\sin\theta$,
and using the Taylor expansion of the exponential function again,
the integral (\ref{intZ2}) becomes 
\begin{eqnarray}
Z&=&
\frac{8\pi^{7/2}}{(2\pi)^{9/2}}\frac{\partial}{\partial \alpha}\;
\sum_{n=0}^{\infty}\frac{\alpha^{2n}}{n!}
W_{n}X_{n}Y_{n},
\end{eqnarray} 
where
\begin{eqnarray}
W_n&=&\int^{\infty}_{0}\!dy\;
\frac{y^{n+1/2}}{(1+y^2)^{n+1/2}}
=\frac{2^{-n-\frac{1}{2}}\sqrt{\pi}\Gamma\left(-\frac{1}{4}+\frac{n}{2}\right)}{\Gamma\left(\frac{1}{4}+\frac{n}{2}\right)}
\qquad (n\ge 1),\\
\nonumber\\
X_n&=&\int^{\pi}_{0}\!d\theta\;
(\sin\theta)^{n-\frac{1}{2}}
=\frac{\sqrt{\pi}\Gamma\left(\frac{1}{4}+\frac{n}{2}\right)}
{\Gamma\left(\frac{3}{4}+\frac{n}{2}\right)},\\
\nonumber\\
Y_n&=&\int^{\infty}_{0}\!dx\;
x^{n+\frac{1}{2}}e^{-\frac{1}{4}x^2}
=2^{n+\frac{1}{2}}\Gamma\left(\frac{3}{4}+\frac{n}{2}\right).
\end{eqnarray}
Since the integral $W_n$ diverges when $n=0$, 
the bosonic partition function ${\cal Z}$ is divergent.
This divergence is, however, independent of $\alpha$.
So Z itself is finite and has the form 
\begin{eqnarray}
Z&=&\frac{1}{\sqrt{2}}\sum_{n=0}^{\infty}\frac{1}{n!}\Gamma\left(\frac{n}{2}+\frac{1}{4}\right)\;\alpha^{2n+1}.
\label{strongZ}
\end{eqnarray}
The formula (\ref{strongZ}) gives us the strong coupling expansion of $Z$
\footnote{We recall that $1/\alpha^2$ is proportional to the coupling constant of NC gauge theory \cite{IKTT}.}.
Furthermore we can carry out the summation in (\ref{strongZ})
as the following
\begin{eqnarray}
\sum_{n=0}^{\infty}\frac{1}{n!}\Gamma\left(\frac{n}{2}+\frac{1}{4}\right)\;\alpha^{2n+1}
\;=\;
\alpha\int^{\infty}_{0}\frac{e^{-z+\alpha^2\sqrt{z}}}{z^{3/4}}dz.
\end{eqnarray}
The $z$-integral can be written by the modified Bessel functions.
Finally we divide $Z$ by the $SU(N)/Z_N$ gauge volume factor:
\begin{eqnarray}
\left.\frac{2^{\frac{N(N+1)}{2}}\pi^{\frac{N-1}{2}}}{2\sqrt{N}\prod_{k=1}^{N-1}k!}\;\right|_{N=2}&=& 2^{\frac{3}{2}}\sqrt{\pi}. 
\end{eqnarray}
In this way, we obtain the final form of $Z$:
\begin{eqnarray}
Z&=&\frac{1}{4\sqrt{\pi}}\;\alpha^2e^{\frac{\alpha^4}{8}}
\Bigg[\pi\sqrt{2}I_{-\frac{1}{4}}\left(\frac{\alpha^4}{8}\right)
-K_{\frac{1}{4}}\left(\frac{\alpha^4}{8}\right)\Bigg],
\label{Z}
\end{eqnarray}
where $I_n(x)$ and $K_n(x)$ are the modified Bessel functions.

\subsubsection*{weak coupling expansion}
In order to compare the above result with the perturbative calculation
of NC gauge theory on fuzzy sphere,
we look for the weak coupling expansion.
From the asymptotic expansion of the Bessel functions:
\begin{eqnarray*}
I_{\nu}(z)&\sim&\frac{e^z}{\sqrt{2\pi z}}
\sum_{n=0}^{\infty}\frac{(-1)^n}{n!}\frac{1}{(2z)^n}\frac{\Gamma(\nu+n+1/2)}{\Gamma(\nu-n+1/2)}
\;+\;\frac{e^{-z+(\nu+1/2)\pi i}}{\sqrt{2\pi z}}
\sum_{n=0}^{\infty}\frac{(-1)^n}{n!}\frac{1}{(2z)^n}\frac{\Gamma(\nu+n+1/2)}{\Gamma(\nu-n+1/2)}\;,\\
\nonumber \\
K_{\nu}&\sim&\sqrt{\frac{\pi}{2z}}e^{-z}
\sum_{n=0}^{\infty}\frac{1}{n!}\frac{1}{(2z)^n}\frac{\Gamma(\nu+n+1/2)}{\Gamma(\nu-n+1/2)}\;,
\end{eqnarray*}
we obtain the weak coupling expansion of the partition function
\begin{eqnarray}
Z&\sim&
\frac{1}{\sqrt{2}}e^{\frac{\alpha^4}{4}}\Bigg[
\sum_{n=0}^{\infty}
\left(\frac{-4}{\alpha^{4}}\right)^n
\frac{1}{n!}\frac{\Gamma(1/4+n)}{\Gamma(1/4-n)}
\Bigg] \nonumber \\
\nonumber \\
&=&\frac{1}{\sqrt{2}}e^{\frac{\alpha^4}{4}}\Bigg[
1+\frac{3}{4}\frac{1}{\alpha^4}
+\frac{165}{32}\frac{1}{\alpha^8}
+\frac{3465}{128}\frac{1}{\alpha^{12}}\;+\;\cdots
\Bigg]
\end{eqnarray}
and the effective action
\begin{eqnarray}
F&=&-\log Z\;\sim\;-\frac{\alpha^4}{4}+\frac{1}{2}\log 2
-\frac{3}{4}\frac{1}{\alpha^4}
-3\frac{1}{\alpha^8}
-\frac{99}{4}\frac{1}{\alpha^{12}}\;+\;\cdots.
\label{weakF}
\end{eqnarray}
On the other hand, the perturbative calculation of the effective action was done up to 2-loop level in \cite{IKTT}.
The 2-loop calculation has the form
\begin{eqnarray}
&&\hspace{-5mm}F_{per}=-\frac{\alpha^4}{6}Nl(l+1)+\frac{1}{2}\log 2
+\frac{1}{\alpha^4}[-n^3(F_{3p}(l)-3F_5(l))-n(F_{3np}(l)-3F_5(l))]
+\;(\mbox{higher loops}).
\nonumber \\
&&\label{pertF}
\end{eqnarray}
$N=n(2l+1)$ is the matrix size. 
In this case $N=2$, so we set $n=1$ and $l=\frac{1}{2}$.
The first term in (\ref{pertF}), 
corresponding to the tree level contribution,
is $-\frac{\alpha^4}{4}$ for $N=2, l=\frac{1}{2}$.
The second term is the one loop contribution.
The subsequent terms are contributions from the two loop calculation.
The definitions of $F_x(l)$ are given in \cite{IKTT}, 
and in our case
$F_{3p}(\frac{1}{2})=F_{3np}(\frac{1}{2})=\frac{3}{8}$.
In this way we confirm that the perturbative calculation reproduces the weak coupling expansion (\ref{weakF}):
\begin{eqnarray}
F_{per}&=&-\frac{\alpha^4}{4}+\frac{1}{2}\log 2-\frac{3}{4}\frac{1}{\alpha^4}\;+\;\mbox{(higher loops)}.
\end{eqnarray}

\section{2-point Function}
We subsequently calculate the normalized expectation values of the gauge invariant operators.
First, we consider the quadratic operator: $\mbox{Tr}A_{\mu}A_{\mu}$.
This operator has a geometrical interpretation: the extension in NC space.  
Due to SO(3) rotational and SU(2) gauge invariance, 
\begin{eqnarray}
\langle\;\mbox{Tr}A_{\mu}A_{\mu}\;\rangle&=&
\frac{3}{2}\;\langle\;(A^1_1)^2\;\rangle\;=\;\frac{3}{2}\langle\;L^2\;\rangle.
\label{AA}
\end{eqnarray} 
The calculation of (\ref{AA}) can be done by the same way as the partition function.
However, an infrared problem occurs; 
$\langle L^2\rangle$ is divergent.
We need to regularize this quantity by adding a mass term $\delta\mbox{Tr}AA$ into the action.
In this regularization, 
$\langle L^2\rangle$
for sufficiently small $\delta$ assumes the following form
\begin{eqnarray}
\langle L^2\rangle&=&
\frac{\gamma+2\int^{\infty}_{0}(e^{\alpha^2\sqrt{z}}-1)e^{-z}\left(\frac{1}{z^{1/4}}+\frac{1}{z^{5/4}}\right)dz}{\int^{\infty}_{0}\frac{e^{\alpha^2\sqrt{z}-z}}{z^{3/4}}dz}
\end{eqnarray}
where $\gamma=\frac{1}{\sqrt{\delta}}\Gamma\left(\frac{1}{2}\right)$.
Using the saddle point method, we can show 
\begin{eqnarray}
\langle\;L^2\;\rangle&\sim&
2^{-\frac{3}{2}}\alpha^3\gamma\;e^{-\frac{\alpha^4}{4}}
+\alpha^2+\frac{4}{\alpha^2}.
\end{eqnarray}
This approximation is valid
for large $\alpha$, i.e. in weak coupling region, where $\langle L^2\rangle\sim\alpha^2$.
It is consistent with the classical behaviour,
since $\frac{1}{N}\mbox{Tr}A^2=\alpha^2l(l+1)$ in a fuzzy sphere solution. 
The $\alpha^{-2}$ behaviour corresponds to the one-loop contribution of NC gauge theory on fuzzy sphere \cite{IKTT2}.
The $\gamma$-dependent part is small for large $\alpha$.
The factor $e^{-\frac{\alpha^4}{4}}$ in the $\gamma$-dependent part implies that it can be identified with the contribution from commutative configurations s.t $[A,A]=0$.
It is because the value of the classical action vanishes for the commutative configurations and equals $-\frac{\alpha^4}{4}$ for the fuzzy sphere configuration.
On the other hand, by using the strong coupling expansion (\ref{strongZ}), we find 
\begin{eqnarray}
\langle\;L^2\;\rangle&=&\frac{\gamma}{\Gamma\left(\frac{1}{4}\right)}+O(\alpha).
\end{eqnarray}
In the strong coupling region, the value of $\langle L^2\rangle$ is 
regularization dependent.
  
\section{3-point Function} 
Next, we consider the expectation value of 
the cubic gauge invariant operator:
$\langle\frac{i}{3}\epsilon_{\mu\nu\rho}\mbox{Tr}A_{\mu}[A_{\nu},A_{\rho}]\rangle$. 
It is easily obtained without divergence as the followings
\begin{eqnarray}
\langle\;\frac{i}{3}\epsilon_{\mu\nu\rho}\mbox{Tr}A_{\mu}[A_{\nu},A_{\rho}]\;\rangle
&=&\frac{\partial F}{\partial \alpha}\;\sim\;
\left\{
\begin{array}{c}
-\alpha^3\qquad \mbox{weak coupling,}\\
\\
-\frac{1}{\alpha} \qquad \mbox{strong coupling.}
\end{array}
\right.
\end{eqnarray}

\section{Summary and Discussion}
In this paper we have carried out an exact calculation of 3-dimensional $N\!\!=\!\!2$ matrix model with Myers term.
A finite partition function has been obtained.
It is interesting also from the point of view of convergence theorem for YM matrix integral \cite{refcn1}\cite{refcn2}. 
Without Myers term, 3d-matrix model does not satisfy the convergence condition.
In our case, however, the introduction of Myers term breaks the basic assumptions for the proof of the theorem \cite{refcn1}, so there is no contradiction between these results.

The weak coupling expansion of the exact effective action reproduces the perturbative calculation in NC gauge theory on fuzzy sphere.
On the other hand, the strong coupling expansion (\ref{strongZ}) shows $F\sim-\log\alpha$.
It is consistent with the conjecture in \cite{IKTT2} and makes our expectation more plausible concerning general behaviour of the effective action in strong coupling region.   
Next, we have calculated the expectation values of the 2-point function: $\mbox{Tr}AA$.
This is a divergent quantity and we need a mass term regularization.
We find the classical and one-loop contributions discussed in \cite{IKTT2} from the regularization independent parts.
The regularization dependent part cannot be identified as the perturbative expansion around the fuzzy sphere configuration.
It should comes from the commutative configuration.
We have also calculated the expectation values of the 3-point function: $\mbox{Tr}\epsilon AAA$.
It requires no regularization but diverges at $\alpha=0$.
The divergence is caused by the singular behaviour of the partition function $Z$ as it vanishes at $\alpha=0$.
However we could expect that $Z|_{\alpha=0}\neq 0$ for generic $N$. 

The extension of NC space is measured by these operators.
It corresponds to the radius of fuzzy sphere in weak coupling region 
where NC space is described classically. 
On the other hand the extension is different from the classical one in strong coupling region.
The infrared divergence gives an infinite value for the range of extension in strong coupling region. 
On the other hand, for $N>2$ or higher dimensional matrix models, the expectation value of $\mbox{Tr}AA$ may be convergent.
The finite extension is expected in this case.
The characterization of the NC space in this region is an interesting problem.  
It is conjectured that 
fuzzy sphere is connected to blanched polymers in strong coupling region \cite{IKTT2}.
We need further information of larger $N$ and higher dimensional models to confirm it.\\

For $N>2$ case and I{}IB matrix model with Myers term,
performing exact calculations  by using the method in this paper seems to be difficult.
The approach from numerical simulations or cohomorogical field theory \cite{MNS} may be useful in those cases. 

\begin{center}
{\bf Acknowledgments}
\end{center}
I would like to thank Y. Kitazawa for discussions and many comments.
I am also grateful to T. Imai and Y. Takayama for discussions.

\def\refname{References}

\end{document}